\documentstyle[twocolumn,aps]{revtex}
\begin{document}
\wideabs{
\draft
\title{Thermodynamics, strange quark matter, and strange stars}
\author{G. X. Peng,$^{1,2,3}$\ H. C. Chiang,$^{2}$\ P. Z. Ning,$^{3}$}
\address{
         $^1$China Center of Advanced Science and Technology
               (World Laboratory), Beijing 100080, China \\
         $^2$Institute of High Energy Physics,
               Chinese Academy of Sciences, Beijing 100039, China \\
         $^3$Department of Physics, Nankai University,
                                Tianjin 300071, China \\
         }
\maketitle
\begin{abstract}
Because of the mass density-dependence, an extra term should be added
to the expression of pressure. However, it should not appear in that of
energy according to both the general ensemble theory and basic 
thermodynamic principle. We give a detail
derivation of the thermodynamics with density-dependent particle masses. 
With our recently determined quark mass scaling, we study strange quark
matter in this new thermodynamic treatment, which still indicates a 
possible absolute stability as previously found. However, the density
behavior of the sound velocity is opposite to the previous finding, but 
consistent with one of our recent publication. We have also studied the 
structure of strange stars using the obtained equation of state.
\end{abstract}
\pacs{PACS numbers: 12.39.-x, 24.85.+p, 05.70.-a, 97.10-q}
         }

\begin{center}
\section{Introduction}
\end{center}          \label{intro}

During the 10-plus years which have elapsed since Witten's conjecture 
\cite{Witten} that strange quark matter (SQM), rather than the normal
nuclear matter, might be the true ground state of quantum chromodynamics
(QCD), many theoretical and observational efforts have been made on the
investigation of its properties and potential astrophysical significance 
\cite{PengCCAST}. 
Because of the well-known difficulty of QCD in the 
nonperturbative domain, phenomenological models reflecting the characteristic
of strong interactions are widely used in the study of hadron, and many of them
have been successfully applied to investigating the stability and 
properties of SQM. One of the most famous models is the MIT bag model
with which Jaffe et.\ al \cite{Jaffe} find that SQM is
absolutely stable around the normal nuclear density for a wide range
of parameters. A vast number of further investigations 
\cite{Madsen,Parija,Sat,Schaffner-Bielich} are performed with fruitful
results. A recent important result is that young millisecond pulsars
are most likely to be strange stars rather than neutron stars
\cite{MadsenPRL98}.
Another alternative model is the mass-density-dependent model with which
Chakrabarty {\it et al}.\ obtained significantly
different results \cite{Cha1,Cha2}. However, Benvenuto and Lugones \cite{Ben}
pointed out that it is caused by the wrong thermodynamic treatment.
They added an extra term to the expression of both pressure and energy,
and got similar results as that in the bag model. 
A recent investigation indicates a link of SQM to the study of quark
condensates \cite{PengPRC56} while a more recent work has carefully studied
the relation between the charge and critical density of SQM
\cite{PengPRC59}.

Latterly, we have demonstrated that the previous treatments 
have unreasonable vacuum limits \cite{PengPRC61}. In addition
to this problem, there exist another serious problem, i.e., the
zero pressure does not appear in the lowest energy state.
In fact, there are two important problems in the quark 
mass-density-dependent model.
One is how to determine the quark mass scaling. The other is how
to deal with the thermodynamics with density-dependent particle
masses self consistently. We have mainly concentrated on the first problem 
in Ref.\ \cite{PengPRC61}. The present paper will concentrate more on 
the second problem. We find that the extra term provided in 
Ref.\ \cite{Ben} should indeed be appended to the expression of pressure.
However, it should not appear in that of energy according to both the
general ensemble theory and basic thermodynamic principle. 
After our modification, the zero pressure point appears exactly at
the lowest energy state, and thus
the thermodynamics with density-dependent particle masses becomes
self-consistent, which leads to completely different density 
behaviour of the sound velocity in SQM and different structure of 
strange stars.

We organize this paper as such. In the subsequent section, we 
give detail arguments on why the additional term in the pressure
should not appear in the energy. The thermodynamic expression
needed later are all derived carefully in this section.
Then in Sec.\ \ref{sqm}, we apply the new thermodynamic formulas to
investigating strange quark matter, and find that the density
behaviour of the sound velocity is opposite to the previous treatment
but consistent with our recent publication. On application of our
obtained equation of state, we integrate the equations of steller
structure for strange stars in Sec.\ \ref{star}, 
and find that strange stars are dimensionally smaller and less massive
than the previous calculation if SQM is absolutely stable. 
Sec.\ \ref{summary} is a short summary.

\begin{center}
\section{Thermodynamics with density-dependent particle masses}
\label{thermo}
\end{center}          

Let us explore directly from the general ensemble theory
what the expression of pressure and energy should look like
if the particle masses are dependent on density.

We express the density matrix as
\begin{equation}
\rho=\frac{1}{\Xi}e^{-\beta \left(E_{N_i,\alpha}-\sum_i\mu_i N_i\right)},
\end{equation}
where $\Xi$\ is the partition function, $\beta$\ is the 
reverse temperature, $N_i$ are the particle numbers,
and $\mu_i$ are the corresponding chemical potentials. 
The microscopic energy $E_{N_i,\alpha}$ is a function of the system volume $V$, 
the particle masses $m_i$,
the particle numbers $N_i$, and the other quantum numbers $\alpha$.
The pressure of the system is 
\begin{eqnarray}
 P&=&\frac{1}{\Xi}\sum\limits_{\{N_i\},\alpha} 
                                   \left(
                      -\frac{\partial E_{N_i,\alpha}}{\partial V}
                                   \right)
                e^{-\beta \left(E_{N_i,\alpha}-\sum_i\mu_i N_i\right)}
                                                                \nonumber  \\
       &=& \frac{1}{\Xi}\sum\limits_{\{N_i\},\alpha}  
                               \left[
                      \frac{1}{\beta}\frac{\partial}{\partial V}
                e^{-\beta \left(E_{N_i,\alpha}-\sum_i\mu_i N_i\right)} 
                              \right]                           \nonumber \\
 &=& \left.\frac{1}{\beta}\frac{\partial\ln\Xi}{\partial V}\right|_{T,\{\mu_i\}}
           = -\left.\frac{\partial(V\Omega)}{\partial V}\right|_{T,\{\mu_i\}},           
\end{eqnarray}
where 
\begin{equation}
 \Omega \equiv -\frac{1}{V\beta}\ln\Xi
\end{equation}
is the thermodynamic potential density which is generally a function of
the temperature $T$, the chemical potentials $\mu_i$, and the particle masses
$m_i$.
If the particle masses have nothing to do with the baryon number density 
$n_b=N/(3V)$ (N is the total particle number), we simply get
\begin{equation}
P=-\Omega.      \label{Pmindep}
\end{equation}
If the masses depend on density or volume, one should have
\begin{equation}  \label{Pgen}
P = -\Omega + n_b\frac{\partial\Omega}{\partial n_b}.
\end{equation}
This is just the right thing the authors have done in Ref.\ \cite{Ben}.
The authors derive it as such:
\begin{equation}
P=\left.-\frac{\partial(\Omega/n_b)}{\partial(1/n_b)}\right|_{T,\mu_i}
 =n_b\frac{\partial\Omega}{\partial n_b}-\Omega.
\end{equation}
 For canonical ensemble, the particle numbers $N_i$ keep fixed. 
 This derivation is thus obvious. However, it is not so obvious for
 grand canonical ensemble because the particle number is not 
 necessarily constant when the temperature T and chemical potentials
 $\mu_i$ unchanged. We will give a more convincing derivation a little later.

The additional term is of crucial importance for pressure balance.
Not as done in Ref.\ \cite{Ben}, however, the extra term does not
appear in the expression of energy.
Now, let's calculate the statistic average for the energy:
\begin{eqnarray}
\bar{E} &=& \frac{1}{\Xi}\sum\limits_{\{N_i\},\alpha}E_{N_i,\alpha}
                   e^{-\beta \left(E_{N_i,\alpha}-\sum_i\mu_i N_i\right)}
                                                       \nonumber \\
        &=& \frac{1}{\Xi}\sum\limits_{\{N_i\},\alpha}\left(
                     -\frac{\partial}{\partial\beta}+\sum_i\mu_i N_i
                                                 \right)
                  e^{-\beta \left(E_{N_i,\alpha}-\sum_i\mu_i N_i\right)}
                                                       \nonumber \\
        &=& -\frac{\partial}{\partial\beta}\ln\Xi+\sum_i\mu_i \bar{N_i},
\end{eqnarray}
where 
\begin{eqnarray}
\bar{N_i}&=&\frac{1}{\Xi}\sum\limits_{\{N_i\},\alpha}  N_i 
               e^{-\beta \left(E_{N_i,\alpha}-\sum_i\mu_i N_i\right)}    
                                                              \nonumber  \\
       &=& \frac{1}{\Xi}\sum\limits_{\{N_i\},\alpha}\left[
                   \frac{1}{\beta} \frac{\partial}{\partial\mu_i}
                   e^{-\beta \left(E_{N_i,\alpha}-\sum_i\mu_i N_i\right)} 
                                                \right]_{V,T,\{m_k\}}
						\nonumber \\
       &=& \frac{1}{\beta}\frac{\partial}{\partial\mu_i}\ln\Xi
            =-V \left.\frac{\partial\Omega}{\partial\mu_i}
                \right|_{T,\{m_k\}}
\end{eqnarray}
is the average number for particle type $i$. Therefore, the energy density
of the system is
\begin{eqnarray}
E&=&\frac{\bar{E}}{V} 
   = \frac{\partial(\beta\Omega)}{\partial\beta}
             +\sum_i\mu_i n_i                              \\
   &=& \Omega+\beta\frac{\partial \Omega}{\partial\beta}
             +\sum_i\mu_i n_i                              \\
   &=& \Omega+\sum_i\mu_i n_i-T\frac{\partial \Omega}{\partial T},
                                            \label{Egen}
\end{eqnarray}
where $n_i$ is the number density of particle type $i$:
\begin{equation}  \label{nigen}
 n_i \equiv \frac{\bar{N_i}}{V}
 =- \left.\frac{\partial\Omega}{\partial\mu_i}
    \right|_{T,\{m_k\}}.
\end{equation}
It is clear from Eq.\ (\ref{Egen}) that only when Eq.\ (\ref{Pmindep}) 
holds can one get the Eq.\ (8) in Ref.\ \cite{Ben}. Therefore, we should not, 
as done in Ref.\ \cite{Ben}, use that expression to calculate the energy density.
Instead, we will calculate $E$ directly from Eq.\ (\ref{Egen}) in this paper.

For more evident arguments, let's see the following derivation starting from
the basic derivative relation for an open system:
\begin{equation} \label{de}
d(VE)= T d(VS) - P dV + \sum_i \mu_i d\bar{N_i},
\end{equation}
where S is the entropy density of the system.

Choosing T, V, and \{$\bar{N_i}$\} as the independent macroscopic state variables, 
the combined statement of the first and second laws of 
thermodynamics, Eq.\ (\ref{de}), can be expressed as
\begin{equation}
d(VA) = -VSdT - P dV + \sum_i \mu_i d\bar{N_i},
\end{equation}
where $A \equiv E-TS$ is the Helmholtz free energy density by which we have
\begin{eqnarray}
P &=& -\left.\frac{d(VA)}{dV}\right|_{T,\{\bar{N_i}\}}                 \\
  &=& -A-V\left.\frac{dA}{dV}\right|_{T,\{\bar{N_i}\}}                \\
  &=& -A+\sum_j n_j \left.\frac{dA(T,\{n_i\})}{dn_j}\right|_{T}.  \label{Pexp3}  
\end{eqnarray}
This is a general expression for pressure. In obtaining the third equality,
we have used the chain relation
\begin{equation}
-V\left.\frac{d}{dV} f\left(\{n_i=\frac{\bar{N_i}}{V}\}
                      \right) 
  \right|_{\{\bar{N_i}\}}
=\sum_j n_j \frac{d}{dn_j} f(\{n_i\}),
\end{equation}
where $f$ is an arbitrary function.

According to the basic relation between thermodynamics and statistics,
we have 
\begin{equation} \label{Adef}
A=\Omega+\sum_i \mu_i n_i
\end{equation}
where $\Omega$\ is the thermodynamic potential density. For a free
Fermi system, it is
\begin{eqnarray}
\Omega &=& -\sum_i \frac{g_iT}{2\pi^2} \int^\infty_0 
\ln\left[1+e^{-\beta\left(\sqrt{p^2+m_i^2}-\mu_i\right)}\right]p^2dp 
                                          \label{Omega0_ori}  \\
&\equiv& \sum_i\Omega_i(T,\mu_i,m_i),
\end{eqnarray}
where $g_i$ is the degeneracy factor which is 6 for quarks and
2 for electrons. In order to include the interaction between
particles, we regard the particle masses as density-dependent, namely
\begin{equation}
m_i=m_i\left(n_b\equiv\sum_jn_j/3\right).
\end{equation}
Because we have chosen $T$, $V$, and 
$\{\bar{N_i}\}$ as independent state variables, the chemical potential 
$\mu_i$ should also be regarded as a function of T and $\{n_k\}$, namely
\begin{equation}
\mu_i=\mu_i\left(T,\{n_k\}\right).
\end{equation}
So, the total
derivative of $\Omega(T,\{\mu_k\},\{m_k\})$ with respect to $n_j$ 
should be taken like this:
\begin{eqnarray}
\left.\frac{d\Omega}{dn_j}\right|_{T} 
&=&\sum_i \left.\frac{\partial\Omega}{\partial\mu_i}\right|_{T,\{m_k\}} 
          \left.\frac{d\mu_i}{dn_j}\right|_T
+\left.\frac{\partial\Omega}{\partial n_b}\right|_{T,\{\mu_k\}}
   \frac{\partial n_b}{\partial n_j}   \\
&=& -\sum_i n_i \left.\frac{d\mu_i}{dn_j}\right|_T
 + \frac{1}{3} \left.\frac{\partial\Omega}{\partial n_b}\right|_{T,\{\mu_k\}}.
\end{eqnarray}
Here we have used the Eq.\ (\ref{nigen}) and the fact that 
${\partial n_b}/{\partial n_j}=1/3$.

Substituting Eq.\ (\ref{Adef}) into Eq.\ (\ref{Pexp3}) gives
\begin{eqnarray}
P &=& -A+  \sum_j n_j \left.\frac{d}{dn_j}
       \left[\Omega +\sum_i \mu_i n_i\right]\right._{T}   \\
  &=& -A + \sum_j n_j 
       \left[\left.\frac{d\Omega}{dn_j}\right|_{T}
  +\sum_i\left(n_i\left.\frac{d\mu_i}{dn_j}\right|_{T} + \mu_i\frac{dn_i}{dn_j} 
         \right) \right]              \\
  &=& -A+\sum_i \mu_in_i+\sum_j \frac{n_j}{3}
                \left.\frac{\partial\Omega}{\partial n_b}\right|_{T,\{\mu_k\}}
                                 \label{Aexp3} \\
&=& -\Omega + n_b \left.\frac{\partial\Omega}{\partial n_b}\right|_{T,\{\mu_k\}} \\
&=& \sum_i\left( 
      -\Omega_i+n_b \frac{\partial m_i}{\partial n_b} 
      \frac{\partial\Omega_i}{\partial m_i}
         \right).   \label{Pfin}
\end{eqnarray}

At zero temperature, the corresponding thermodynamic potential density 
can be obtained from Eq.\ (\ref{Omega0_ori}) by first
taking the limit $T\rightarrow{0}$ and then carrying out the resulting
integration:
\begin{eqnarray}                   \label{Omegai}
 & \Omega= -\sum\limits_i \frac{g_{i}}{48\pi^{2}}
\biggl[\mu_{i}(\mu_{i}^{2}-m_{i}^{2})^{1/2}(2\mu_{i}^{2}-5m_{i}^{2})&
                                              \nonumber   \\
  &+3m_{i}^{4}\ln\frac{\mu_{i}
         +\sqrt{\mu_{i}^{2}-m_{i}^{2}}}{m_{i}}\biggr].&
 \end{eqnarray}
We thus have from Eqs.\ (\ref{nigen}), (\ref{Egen}), and (\ref{Pfin}):
\begin{eqnarray}
n_{i} &=& \frac{g_i}{6\pi^2}(\mu_i^2-m_i^2)^{3/2}, \label{n0} \\
E  &=& \sum\limits_{i} m_i n_i F(x_i),       \label{E0}   \\
 P  &=& \sum\limits_{i} m_i n_i x_i^2 G(x_i)
       -\sum\limits_{i} m_i n_i f(x_i)       \label{P0} 
\end{eqnarray}
where
\begin{equation}
 x_i \equiv \frac{p_{f,i}}{m_i}
 =\frac{\left(\frac{6\pi^2}{g_i}n_{i}\right)^{1/3}}{m_i}
 =\frac{\sqrt{\mu_i^2-m_i^2}}{m_i}
\end{equation}
is the ratio of the Fermi momentum to the mass that related to
particle type $i$. With the hyperbolic sine function
sh$^{-1}(x)\equiv\ln(x+\sqrt{x^2+1})$, the functions $F(x_i)$, $G(x_i)$, and
$f(x_i)$ are defined as
\begin{eqnarray}
 & F(x_i) \equiv \frac{3}{8}
    \left[
          x_i \sqrt{x_i^2+1}(2x_i^2+1)-\mbox{sh}^{-1}(x_i)
    \right]/x_i^3,  &   \\
 & G(x_i) \equiv \frac{1}{8}
    \left[
          x_i\sqrt{x_i^2+1}(2x_i^2-3)+3\mbox{sh}^{-1}(x_i)
    \right]/x_i^5,  &    \\
 &f(x_i) \equiv -\frac{3}{2}\frac{n_b}{m_i}\frac{dm_i}{dn_b}
                  \left[
               x_i \sqrt{x_i^2+1}-\mbox{sh}^{-1}(x_i)
                  \right]/x_i^3.  &
\end{eqnarray}

   One can see, from Eqs.\ (\ref{E0}) and (\ref{P0}), that an additional
term appears in the pressure expression, but not in the energy expression.
We can specially confirm this result further as such.

From Eq.\ (\ref{de}), one has an alternative general expression
for pressure:
\begin{eqnarray} 
P &=& -\left.\frac{d(VE)}{dV}\right|_{S,\{\bar{N_k}\}}  \label{Pgen2} \\
 &=& -E+ \sum_j n_j \left.\frac{dE}{dn_j}\right|_{S}. 
                            \label{Pde}
\end{eqnarray}

According to the Pauli principle and the relativistic energy-momentum
relation $\varepsilon_i=\sqrt{p^2+m_i^2}$, the energy density of the system
at zero temperature should be
\begin{equation}
  E(\{n_{i}\},\{m_j(n_b)\}) = \sum\limits_i \frac{g_i}{2\pi^2}
   \int^{p_{f,i}}_0 \varepsilon_i p^2 dp,   \label{Edef}
\end{equation}
which, when the integration is carried out, 
is just the same as Eq.\ (\ref{E0}).

Because the entropy is also zero (or constant) at zero temperature,
we can substitute Eq.\ (\ref{Edef}) into Eq.\ (\ref{Pde}), and accordingly get
\begin{eqnarray}
P &=&-E+\sum_j n_j \left( \frac{\partial E}{\partial n_j}
           +\sum_i \frac{\partial E}{\partial m_i} 
     \frac{\partial m_i}{\partial n_b} \frac{\partial n_b}{\partial n_j}
                     \right)                 \\
   &=& -E + \sum_j n_j \frac{\partial E}{\partial n_j}
     + \sum_i \sum_j \frac{n_j}{3}\frac{\partial E}{\partial m_i} 
                \frac{\partial m_i}{\partial n_b}   \\
   &=& -\Omega + \sum_i n_b\frac{\partial m_i}{\partial n_b}
                  \frac{\partial E}{\partial m_i},
\end{eqnarray}
which leads to Eq.\ (\ref{P0}) exactly.

\begin{center}
\section{Properties of strange quark matter in the new thermodynamic treatment}
 \label{sqm}
\end{center}         

Having derived in detail the thermodynamics with variable particle masses in
the previous section, we now apply it to the investigation of strange quark
matter.

  As usually done in the previous literature 
\cite{Jaffe,Cha1,Cha2,Ben,PengPRC56,PengPRC59,PengPRC61}, 
We assume the SQM to be a Fermi gas
mixture of $u$, $d$, $s$ quarks and electrons with chemical 
equilibrium maintained by the weak interactions:
$$
 d, s \leftrightarrow u+e+\overline{\nu}_{e}, \, \,
                     s+u \leftrightarrow u+d, ...
$$
Because of these reactions, the chemical potentials $\mu_i\ (i=u, d, s, e)$
should satisfy
\begin{eqnarray}
   &  \mu_d  = \mu_s \equiv \mu,  &    \label{eqmu1}     \\
   & \mu_u + \mu_e = \mu.         &    \label{eqmu2}                
\end{eqnarray}

For the bulk SQM in weak equilibrium, the previous
investigations got a slightly positive charge \cite{Jaffe}. Our recent study
\cite{PengPRC59} demonstrates that negative charges could lower the critical
density. However, too much negative charge can make it
impossible to maintain flavor equilibrium. Therefore, the charge of SQM is
not allowed to shift too far away from zero at both positive and
negative directions. Therefore, one also has another two equations
for a given baryon number density $n_b$:
\begin{eqnarray}
   & \frac{1}{3} (n_u + n_d + n_s) = n_b, &     \label{eqmu3}         \\
   & \frac{2}{3}n_u-\frac{1}{3}n_d-\frac{1}{3}n_s-n_e = 0.
                                      \label{eqmu4}   &
\end{eqnarray}
The first is the definition of baryon number density; the second
is from the charge neutrality requirement. $n_i\ (i=u, d, s, e)$
is related to $\mu_i$ and $m_i$ by Eq.\ (\ref{n0}).

Because the results from lattice calculations show that quark matter
does not become asymptotically free soon after the phase 
transition (instead, it approaches the free gas equation of state
very slowly), one should consider the strong interaction between
quarks in a proper way. We do this by including the interaction effect
within the variable quark masses. Because of the characteristic of 
the quark confinement and asymptotic freedom of the strong interaction, 
one can write down the simplest and most symmetric
parametrization for the quark masses $m_q\ (q=u, d, s)$ \cite{PengPRC61}:
\begin{equation} \label{mqji}
m_q= m_{q0}+\frac{D}{n_b^z},
\end{equation}
where $m_{q0}$ is the corresponding quark current mass, 
$z$ is a fixed exponential. Previously, $z$ is regarded as 1. 
Our recent study \cite{PengPRC61} indicates that it is more reasonable 
to take $z=1/3$. 
The parameter $D$ is usually determined by stability arguments,
 i.e., at zero pressure $(P=0)$, the energy per baryon, $E/n_b$, is great
 than 930 MeV for two flavor quark matter in order not to contradict 
 standard nuclear physics, but less than 930 MeV for three flavor symmetric
 quark matter so that SQM can have the possibility of absolute stability.
Obviously, the rang of $D$ determined by this method depends on 
different thermodynamic treatments. Within the thermodynamics derived
in the preceding section, $D$ is in the range (155---171 MeV)$^2$ 
when taking $z=1/3$. 

Because the light quark current masses are very small, their value
uncertainties are not important. So we take the fixed central
values $m_{u0}=5$ MeV and $m_{d0}=10$ MeV in our calculation. The
electron mass is very small (0.511 MeV). As for $s$ quarks, 
we take 80 and 90 MeV, corresponding respectively
to $D^{1/2}$ = 156 and 160 MeV.

For a given $n_b$, we solve for $\mu_i\ (i=u,d,s,e)$ from
Eqs.\ (\ref{eqmu1})---(\ref{eqmu4}), 
and calculate the energy density and pressure of SQM respectively from 
Eqs.\ (\ref{E0}) and (\ref{P0}) with the quark messes replace by
Eq.\ (\ref{mqji}).

Firstly, we draw the configuration of the SQM for the parameter set
$m_{s0}=80$ MeV and $D^{1/2}=156$ MeV in Fig.\ \ref{frac}.
At high densities, all of the $u, d$, and $s$ quarks tend to become
a triplicate. When the density becomes lower, $d$ fraction will increases
while $s$ fraction decreases, and will become zero 
at a definite density which is called critical density in Ref.\ 
\cite{PengPRC59} because SQM does not exist below that density.  
The $u$ fraction is nearly unchanged. 
It in fact increases very slowly. To keep charge neutrality, the electron
fraction will also increase. However, because of its very small mass, 
the electron fraction is so little that we multiply by one thousand 
to draw it in the figure. 

In Fig.\ \ref{enb}, we show the density dependence of the energy per baryon,
$VE/N=E/n_b$, vs baryon number density $n_b$ for the parameter set
I: $m_{s0}=80$ MeV, $D^{1/2}=156$ MeV, and II:
$m_{s0}=90$ MeV, $D^{1/2}=160$ MeV. For the first parameter set, SQM is
absolutely stable while for the second set it is nearly meta-stable.
The point marked with a circle `$\circ$' is
the zero pressure point where the system pressure becomes zero. 
It can be clearly seen that the zero pressure points are exactly 
located at the lowest energy state. In fact, this is a basic requirement
of thermodynamics because one can obtain from Eq.\ (\ref{Pgen2})
\begin{equation}
  P=-\frac{d(VE)}{dn_b} \left.\frac{dn_b}{dV}\right|_{\{\bar{N_k}\}}
   =-n_b^2 \frac{d(E/n_b)}{dn_b}.
\end{equation}
However, this is not the case for most of the previous thermodynamic
treatments of strange quark matter in the mass-density-dependent model
\cite{Ben,Cha1,Cha2,PengPRC56}, which is their another serious flaw in 
addition to the unreasonable vacuum limits mentioned before. 

In Fig.\ \ref{eos}, we give the relation between the pressure $P$ and energy
density $E$, i.e., the equation of state. It approaches the free gas 
equation of state at high densities. However, its shape is a little sunken 
at lower densities, contrary to previous calculation which is protuberant. 
This will leads to completely different lower density behaviour of the sound 
velocity in strange quark matter.

The velocity of sound is plotted in the lower 
part of Fig.\ \ref{vos}. The upper part is calculated by the same 
method in Ref.\ \cite{Ben} with parameter set B there. Simultaneously
given with a full horizontal line is the ultra-relativistic case 
($1/\sqrt{3}$) for purpose of comparison. Obviously, they become nearly
identical at high densities while the lower density behaviour is opposite.
The sound velocity in the previous treatment is higher than the 
ultra-relativistic case and will eventually exceed the speed of light
at lower densities, which is unreasonable from the point of view of 
the theory of relativity. 

\begin{center}
\section{          Strange stars }
\label{star}
\end{center}

It has long been proposed that the currently called neutron stars might be  
composed of strange quark matter and thus be in fact strange stars.
A recent investigation shows that young millisecond pulsars
are most likely to be strange stars rather than neutron stars
\cite{MadsenPRL98}.
Previous authors have investigated the properties of strange stars
by applying their obtained equation of state with interesting results
\cite{Cha2,Ben}. We have now modified
the thermodynamic treatment and updated the quark mass scaling. Therefore,
it is meaningful to study the structure of strange stars in the 
new context from the astrophysical view point.

As generally done, we assume the strange star to be a spherically 
symmetric object. Its stability is determined by the general relativistic
equation of hydrostatic equilibrium known as Tolman-Oppenheimer-Volkov
equation \cite{Lip}
\begin{equation} \label{TOV}
\frac{dP}{dr}=-\frac{GmE}{r^2}\frac{(1+4\pi r^3P/m)(1+P/E)}{1-2Gm/r},
\end{equation}
with the subsidiary condition
\begin{equation} \label{TOVsub}
dm/dr=4\pi r^2E,
\end{equation}
where $G=6.707*10^{-45}$ MeV$^{-2}$ is the gravitational constant, $r$ is
the distance from the core of the star, $E=E(r)$ is the energy density
or mass density, $P=P(r)$ is the pressure, and $m=m(r)$ is the mass 
within the radius $r$.

For an initial baryon number density $n_0$ (accordingly $P_0$ and $E_0$), 
we can numerically solve Eqs.\ (\ref{TOV}) and (\ref{TOVsub}) with the 
aid of the equation of state, and obtain the corresponding $P=P(r,n_0)$
and $m=m(r,n_0)$, and consequently $n=n(r,n_0)$, the baryon number density
at the radius $r$ for the central density $n_0$. 
The radius $R$ of the strange star is determined by the condition
\begin{equation}
P(R,n_0)=0,
\end{equation}
namely, 
\begin{equation}
R=R(n_0).
\end{equation}
Accordingly, the mass of the strange star is
\begin{equation}
M=m(R(n_0),n_0) \equiv M(n_0).
\end{equation}

To make strange stars stable, we must require $dM/dn_0>0$. 
For the above obtained equation of state, M firstly increases with
$n_0$ up to a definite value $M_{max}$ corresponding to the 
highest acceptable central density $n_{0max}$. After that, $M$ decreases
with $n_0$, and the star becomes unstable.

For the parameter set I, i.e.\ $m_{s0}=80$ MeV and $D=(156$ MeV$)^2$, 
we give the density profiles $n(r,n_0)$ in Fig.\ \ref{profile} as an 
example. The upmost line is for the largest acceptable central density
$n_{0max}$ $(\approx 1.35$ fm$^{-3}$). The lowest horizontal line 
corresponds to the surface density $n_s$ $(\approx 0.25$ fm $^{-3}$) of 
strange stars which is independent of the central density, but a functional
of the equation of state.
Each line will intersect with it. The cross points correspond to 
the radius $R$ of the star. The maximum radius of the star appears
in $n_0$ $\approx 0.65$ fm$^{-3}$.

In Fig.\ \ref{MR}, we show the mass-radius relation of strange stars
with a solid line. The point marked with a full dot `$\bullet$' represents
the largest acceptable mass $M_{max}$ ($\approx 1.58$ times the solar mass).
For comparison, we have also plotted the result from the bag model calculation
with the bag constant $B^{1/4}=144$ MeV, and that in Re.\ \cite{Ben} with 
the parameter set B there. We can see that the strange stars in our case 
is dimensionally smaller and less massive than the previous calculation
if SQM is absolutely stable. Naturally, this observation depends on the 
parameters employed. If we choose a bigger $m_{s0}$ and larger $D$, 
the case would be different. However, SQM would have no possibility of 
absolute stability in that case.

\begin{center}
\section{          Summary}
\label{summary}
\end{center}

We have derived the thermodynamics with density-dependent
particle masses self-consistently, which overcome the serious flaws
of the previous treatment of SQM in the quark mass-density-dependent
model. We find that an additional term should be appended to 
the expression of pressure, but it should not appear in that of energy.
When applying the new formulas to the investigation of SQM,
we find that the density behaviour of the sound velocity is
opposite to the previous calculation, but consistent with our
recent publication, which leads to different structure of strange
stars.

\begin{center}
\section*{          ACKNOWLEDGMENTS }
\end{center}

The authors would like to thank the National Natural Science Foundation
of China for financial support under Grant No.\ 19905011.

\begin{figure}
\caption{The configuration of SQM varies with density. At high densities, 
 all of the $u, d$, and $s$ quarks tends to become a triplicate.
 When the density becomes lower, $d$ fraction will increases while  
 the $s$ fraction decreases. The $u$ fraction is nearly unchanged. 
 It only increases very slowly. 
         }
\label{frac}
\end{figure}

\begin{figure}
\caption{Energy per baryon $E/n_b$ vs baryon number density $n_b$ for 
  differnt parameter groups. The zero pressure points marked with `$\circ$' 
  are located at the lowest energy state, which is not the 
  case for most of the previous thermodynamic treatments of strange quark
  matter in the quark mass-density-dependent model. 
         }
\label{enb}
\end{figure}

\begin{figure}
\caption{Equation of state of strange quark matter (pressure P vs energy density
  E). it approaches the free gas equation of state at high densities. However, it
  is a little sunken at lower densities, contrary to previous calculation.}
\label{eos}
\end{figure}

\begin{figure}
\caption{Velocity of sound in strange quark matter. The solid horizontal line
  is the ultra-relativistic case. the lower half part is the results of our 
  calculation while the upper part is calculated by the same method 
  in Re.\ \protect\cite{Ben} for parameter set B there. 
  Their lower density behaviour is obviously opposite.
     }
\label{vos}
\end{figure}

\begin{figure}
\caption{Density profiles for parameter set $m_{s0}=80$ MeV and 
 $D=(156$ MeV)$^2$. The upmost line is for the largest acceptable 
 central density $n_{0max}$. The lowest horizontal line corresponds
 to the surface density of strange stars. 
 The cross points of each line and the lowest horizontal line 
 correspond to the radius $R$ of the star.
      }
\label{profile}
\end{figure}

\begin{figure}
\caption{Mass-Radius relation for strange stars. The vertical axis
is the star mass in unit of the solar mass while the horizontal axis
is the star radius in unit of kilometer. The solid line
is obtained by the method in this paper. The dotted line is
from the bag model. The dashed line is calculated with the same
method in Ref.\ \protect\cite{Ben} for the parameter set B there. 
The points marked with a full dot `$\bullet$' represent the maximum
acceptable masses.
     }
\label{MR}
\end{figure}

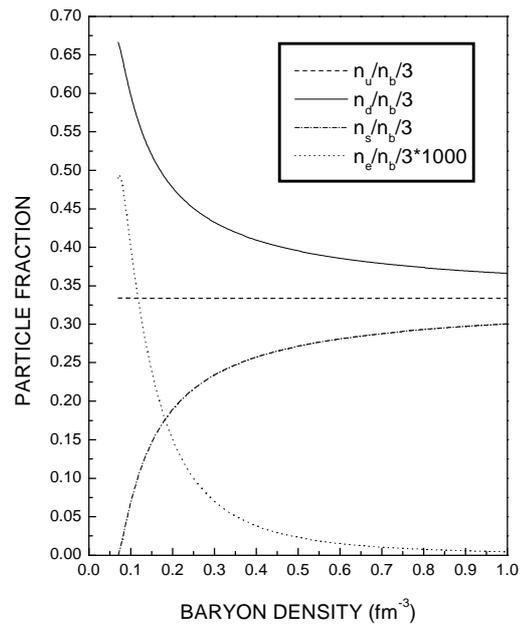
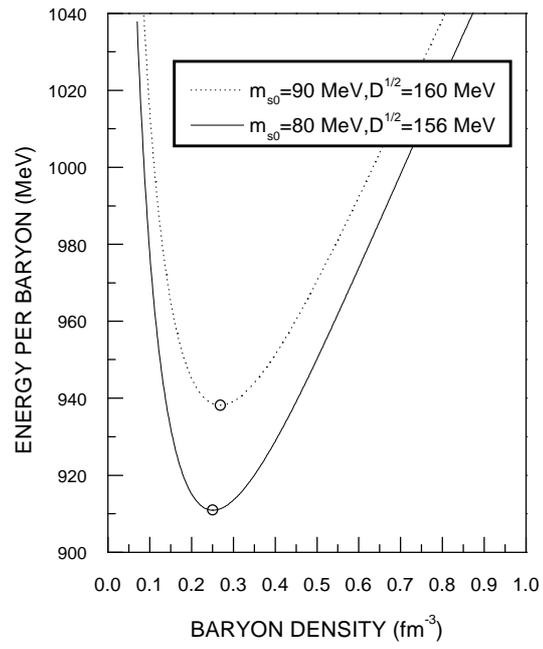
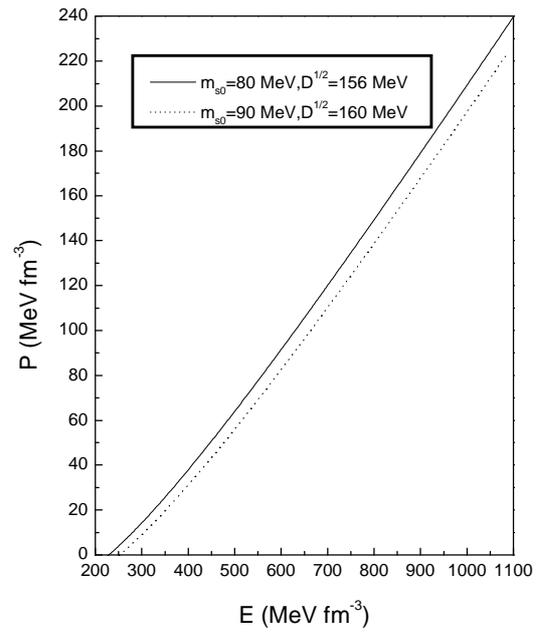
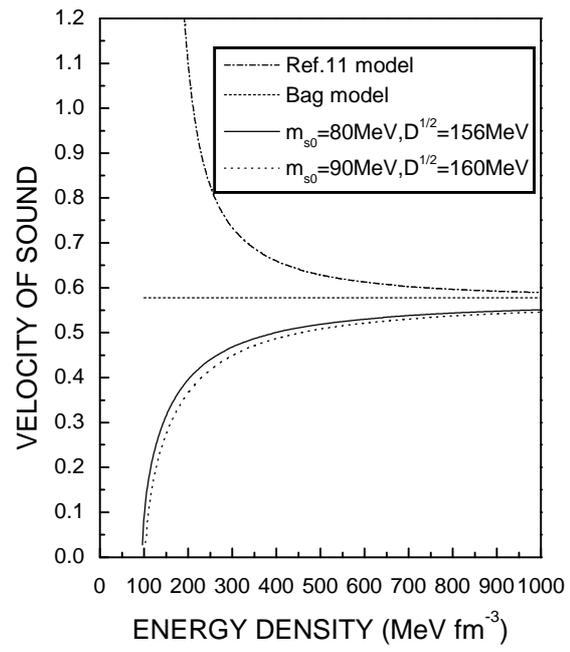
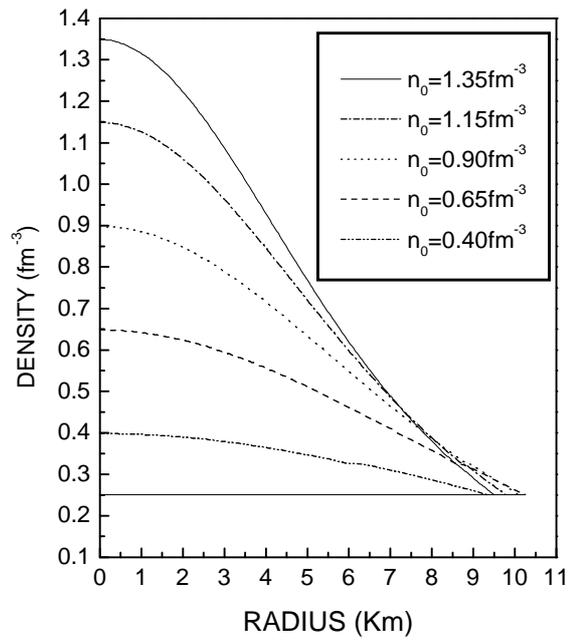
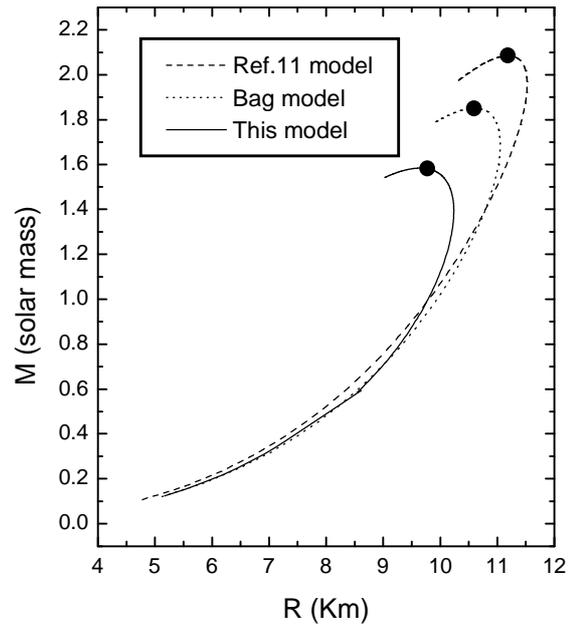
\end{document}